\documentclass{amsart}

\newtheorem{theorem}{Theorem}[section]

\theoremstyle{definition}
\newtheorem{definition}[theorem]{Definition}
\newtheorem{example}[theorem]{Example}

\theoremstyle{remark}

\numberwithin{equation}{section}
\newcommand{\abs}[1]{\lvert#1\rvert}

\def\C{\mathbb{C}}
\def\I{\mathbb{I}}
\def\N{\mathbb{N}}
\def\D{\mathfrak{D}}
\def\H{\mathfrak{H}}
\def\an{\mathfrak{a}}
\def\en{\mathfrak{n}}
\def\WH{\widehat{\mathfrak{H}}}
\def\G{\mathfrak{G}(r_1,r_2)}
\def\g{G(r_1,r_2)}
\begin{document}

\title{Vector coherent states with matrix moment problems}
\author{K. Thirulogasanthar}
\address{Department of Mathematics and Statistics, Concordia
University, 7141 Sherbrooke Street West, Montreal, Quebec H4B 1R6,
Canada} \email{santhar@cs.concordia.ca}
\author{A.L. Hohou\'eto}
%\address{Department of Mathematics and Statistics, Concordia
%University, 7141 Sherbrooke Street West, Montreal, Quebec H4B 1R6,
%Canada }
\email{al\_hohoueto@yahoo.fr}

\begin{abstract}
Canonical coherent states can be written as infinite series in
powers of a single complex number $z$ and a positive integer
$\rho(m)$. The requirement that these states realize a resolution
of the identity typically results in a moment problem, where the
moments form the positive sequence of real numbers
$\{\rho(m)\}_{m=0}^\infty$. In this paper we obtain new classes of
vector coherent states by simultaneously replacing the complex
number $z$ and the moments $\rho(m)$ of the canonical coherent states by $n \times n$ matrices. Associated oscillator algebras are discussed with the aid of a generalized matrix factorial.
Two physical examples are discussed. In the first example coherent states are obtained for the
Jaynes-Cummings model in the weak coupling limit and some physical properties are discussed in terms of the constructed coherent states. In the second example coherent states are obtained  for a conditionally exactly solvable supersymmetric radial harmonic oscillator. 
\end{abstract}

\maketitle

\section{Introduction}
Overcomplete family of vectors of Hilbert spaces play a pivotal role in quantum theories, signal and image analysis. The most fundamental component in the analysis of the states in quantum Hilbert space of a physical problem is an overcomplete family of vectors known as coherent states (CS for short). The wide use of CS in quantum theories and in other scientific areas has developed the theory of CS to a tremendous extent. Connections between CS and group representations, orthogonal polynomials,
Lie algebras have been studied extensively \cite{AAG,KS,P,B,O}.

It is understood that the well-known canonical CS of the harmonic oscillator are described equivalently as eigenstates of the usual bosonic annihilation operator, the trajectory of a displacement operator acting on a fundamental state and as minimum uncertainty states. However, what we observe in different generalizations of the canonical CS is that the preceding equivalence is no longer present. For example, the Barut-Girardello CS are not minimum uncertainty states but they satisfy the other two properties \cite{BG,P}. Moreover, classes of CS were derived as eigenstates of certain operators associated to Hamiltonians \cite{Coo,Po}. Most of the CS given in \cite{KPS} can only be seen as an eigenstate of a generalized annihilation operator. There are several articles in the literature where CS  are obtained by defining them as minimum uncertainty states, for example see \cite{NS}.

 Properties of the harmonic oscillator canonical CS are well-known. In generalizing the definition of the canonical CS we always intend to keep as many properties of the canonical CS as possible. There are a number of generalized definitions for a set of
CS; for different approaches see \cite{AAG,KS,P,GK, NG, TA, AEG}. In this article we follow the following generalization of the canonical CS and generalize it a step further.
%-------
 \begin{definition}\label{def1}
Let $\mathfrak{H}$ be a separable Hilbert space with an
orthonormal basis $\{\phi_m\}_{m=0}^\infty$ and $\C$ be the
complex plane. For $z \in\D$, an open subset of $\C$, the states
 \begin{equation}
\mid z \rangle = N(|z|)^{-1/2} \sum_{m=0}^\infty
\frac{z^m}{\sqrt{\rho(m)}} \phi_m \in  \H
 \label{e1}
 \end{equation}
are said to form a set of coherent states if the following
conditions hold:
 \begin{enumerate}
 \item[(i)] For each $z \in \D$, the state $\mid z \rangle$ is
normalized, that is, $\langle z \mid z \rangle = 1$;
 \item[(ii)] The set, $\{\mid z \rangle : z \in \D\}$ permits a
resolution of the identity, that is,
 \begin{equation}
\int_\D \mid z \rangle  \langle z \mid d\mu = I,
 \end{equation}
 \end{enumerate}
where $N(|z|)$ is a normalization factor,
$\{\rho(m)\}_{m=0}^\infty$ is a positive sequence of real numbers,
and $d\mu$ is an appropriately chosen measure on $\D$.
 \end{definition}
%-------
Vector coherent states are well-known mathematical objects, often they are defined as orbits of vectors under the operators of unitary representations of groups \cite{AAG, RD}. However,
in \cite{TA} vector coherent states (VCS for short) were developed as $n$ component vectors in a Hilbert space $\C^n\otimes \H$
by replacing the complex variable $z$ of (\ref{e1}) by an $n
\times n$ matrix
 \begin{equation}
 \label{ee1}
Z = A(r) e^{i\zeta\Theta(k)},
 \end{equation}
where $A(r)$ and $\Theta(k)$ are $n \times n$ matrices such that
 \begin{equation}
 \label{ee2}
[A(r),A(r)^\dagger] =0, \quad \Theta(k) =
\Theta(k)^\dagger, \quad [A(r),\Theta(k)] = 0,
 \end{equation}
where the variables $r$, $k$ and $\zeta$ live in appropriate
measure spaces, and $M^\dagger$ stands for the transposed complex
conjugate of the matrix $M$. In \cite{AEG} as a further generalization of \cite{TA} VCS were studied as infinite component vectors in a suitable Hilbert space. The term VCS was used in \cite{TA,AEG} to describe that when the complex number $z$ of definition \ref{def1} is replaced by an $n\times n$ square matrix we obtain CS as $n$ component vectors. However, in \cite{TA} for some particular cases the link to a group representation was derived.

The physical motivation of the generalization given in this paper is the construction of CS for multi-level quantum systems with non-degenerate discrete spectrum. In the literature, for two level atoms CS were constructed in the form of (\ref{e1}) to each level \cite{DH}. In the present scheme, using matrices, we develop a more systematic method to derive CS for multi-level quantum systems with non-degenerate infinite energy spectrum. We shall also apply the same method to construct  CS for supersymmetric Hamiltonians with non-degenerate energies. Here again the method is different from the ones appearing in the literature (see Subsection \ref{susy}).

The simplest model in use for the description of a single two-level atom interacting with a single cavity mode of the electromagnetic field is the Jaynes-Cummings model (JC). This model is exactly solvable in the rotating wave approximation, where one may use the diagonalization technique to solve it \cite{DH}. If one neglects losses, multi-mode multi-level generalizations of the JC can be solved exactly using the exact solvability of the JC \cite{Ja,As,Gao}. Suppose we have a diagonalizable Hamiltonian $H$ for a $n$-level atom in a single-mode cavity field with non-degenerate energies $E_m^k$ and wavefunctions $\psi_m^k$, $k=1,2,...,n;~m=0,1,2,...\infty.$ Let $x_m^k=E_m^k-E_0^k;~k=1,...n$. Let $R(m)=[\text{diag}(x_m^1!,...,x_m^n!)]$ and $Z=\text{diag}(z_1,...,z_n)$ be diagonal matrices, where $x_m^k!=x_1^k...x_m^k$ is the generalized factorial. Assume that the following vectors are normalized and satisfy a resolution of the identity,
\begin{equation}\label{mo}
\mid Z,k\rangle=N(Z)^{-\frac{1}{2}}\sum_{m=0}^{\infty}R(m)^{-\frac{1}{2}}Z^m\Psi_m^k;~~k=1,2,...,n
\end{equation}
where $\Psi_m^k:=(0,...0,\psi_m^k,0,...,0),$ and $\psi_m^k$ is placed in the $k$ th position. The collection of vectors (\ref{mo}) forms a set of CS for the diagonalized Hamiltonian $H_D$. A general set of CS for the Hamiltonian $H_D$ can be written as
$$\mid Z\rangle=\sum_{k=1}^{n}c_k\mid Z,k\rangle\;\;\;\text{with}\;\;\;\sum_{k=1}^{n}\abs{c_k}^2=1.$$
Further, if $O$ is the diagonalization operator such that $H=OH_DO^{\dagger}$ then the above sets of CS can be transformed as CS of $H$ with the aid of the operator $O$. A similar argument for a two-level system leading to the quaternionic VCS of \cite{TA} was given in \cite{AEG}.

The states (\ref{mo}) can be considered as a generalization of (\ref{e1}), in which the complex number $z$ and the positive sequence of real numbers $\rho(m)$ are replaced by $n\times n$ diagonal matrices. Motivated from the above discussion, in this article, as a generalization to definition \ref{def1} and to \cite{TA}, we construct
VCS by replacing both the complex variable $z$ and the positive
numbers $\rho(m)$ by $n \times n$ matrices $Z$ and $R(m)$
respectively. To be more general, we will carry out our construction with more general matrices than the diagonal ones. In order to be consistent with one of the three equivalent definitions of the canonical CS we introduce an oscillator algebra by defining a matrix factorial and realize the VCS as eigenstates of a generalized annihilation operator. As a physical example of the construction, the JC model in quantum optics can be taken. We shall justify this claim in Section \ref{phys} and use the constructed VCS to obtain various physical quantities associated to the problem. These quantities may be used to justify the validity of the construction. Apart from quantum mechanical point of view, the following sets of VCS are continuous tight frames and thereby they may find applications in multi-channel signal processing. 

 Since we have replaced $z$ and $\rho(m)$ by
matrices, and matrices do not commute in general, the order in
which the products of matrices are computed is primordial, that
is, we can choose either the order $R(m)Z^m$ or $Z^m R(m)$ (that
we will call from now on {\em ``R-Z ordering"} and {\em ``Z-R
ordering"}, or, abusively, {\em "R-Z representation"} and {\em
``Z-R representation"}). Whatever the choice is, the construction
of VCS ends up with a matrix moment problem. However, a crucial
fact appears: According to the ``representation" used, the
construction of CS may succeed in one case and fail in the other case.

%----------------------------------
\section{VCS with the R-Z ordering}
\label{general}
%-----------------------------------
Let $r \in [0,\infty)$ and $\zeta \in [0,2\pi)$. Let $A(r)$ and
$R(m)$ be $n \times n$ matrices. Set $Z = A(r) e^{i\zeta}$. Let
$\chi^1,\ldots,\chi^n$ be the canonical orthonormal basis of
$\C^n$ and $\{\phi_m\}_{m=0}^\infty$ be an orthonormal basis of an
abstract separable Hilbert space $\H$. Let $\WH = \C^n \otimes
\H$. Then, $\{\chi^j \otimes \phi_m : j=1,\dots,n, \; m \in \N\}$
is an orthonormal basis of $\WH$. Define the set of states
 \begin{equation}
\mid Z,j \rangle = N(|Z|)^{-1/2} \sum_{m=0}^\infty R(m) Z^m \chi^j
\otimes \phi_m \in \WH, \quad j=1,2,\dots,n,
 \label{e2}
 \end{equation}
and denote 
 \begin{equation}
|M| = [MM^\dagger]^{1/2} = [M^\dagger M]^{1/2}.
 \end{equation}
%-------
 \begin{theorem}
 \label{Th1}
The states in (\ref{e2}) are VCS in the sense that they satisfy
the normalization condition and realize a resolution of the
identity, that is,
 \begin{eqnarray}
\label{e3} \sum_{j=1}^n \langle Z,j \mid Z,j \rangle &=& 1, \\
\label{e4} \int_{0}^\infty \int_0^{2\pi} \sum_{j=1}^n \mid Z,j
\rangle \langle Z,j \mid W(|Z|) d\mu &=& \I_n \otimes I,
 \end{eqnarray}
provided that
 \begin{equation}
 \label{e5}
N(|Z|) = \sum_{m=0}^\infty {\text{Tr}} \{ [R(m)A(r)^m]^\dagger
[R(m)A(r)^m] \} = \sum_{m=0}^\infty {\text{Tr}} ~|R(m)A(r)^m|^2 <
\infty,
\end{equation}
and
 \begin{equation}
 \label{ee6}
2\pi \int_0^\infty N(|Z|)^{-1} [R(m)A(r)^m] [R(m)A(r)^m]^\dagger
W(|Z|) d\nu = \I_n,
 \end{equation}
where $d\nu$ and $d\mu$ are  appropriate measures on $[0,\infty)$
and $[0,\infty) \times [0,2\pi)$ respectively, and $W(|Z|)$ is a
positive weight function.
 \end{theorem}
%-------
%-------
 \begin{proof}
We have that
 \begin{eqnarray*}
\sum_{j=0}^n \langle Z,j \mid Z,j \rangle &=& N(|Z|)^{-1}
\sum_{j=0}^n \sum_{m=0}^\infty \sum_{l=0}^\infty \langle R(m)Z^m
\chi^j \mid R(l)Z^l \chi^j \rangle_{\C^n} \langle \phi_m
\mid \phi_l \rangle_{\H} \\
&=& N(|Z|)^{-1} \sum_{m=0}^\infty \text{Tr} |R(m)A(r)^m|^2 = 1,
 \end{eqnarray*}
On the other hand, for $d\mu = d\nu d\zeta$, we have:
 \begin{eqnarray*}
\lefteqn{ \int_0^\infty \int_0^{2\pi} \sum_{j=1}^n \mid Z,j
\rangle \langle Z,j \mid W(|Z|) d\mu = {} } \\
&=& {} \sum_{j=1}^n \sum_{m=0}^\infty \sum_{l=0}^\infty
\int_0^\infty \int_0^{2\pi} N(|Z|)^{-1} \mid R(m)Z^m \chi^j
\otimes \phi_m \rangle {} \\
& & {} \hspace{5cm} \langle R(l)Z^l \chi^j \otimes \phi_l
\mid W(|Z|) d\mu {} \\
&=& {} \sum_{m=0}^\infty \sum_{l=0}^\infty \int_0^\infty
\int_0^{2\pi} N(|Z|)^{-1} [R(m)Z^m] \left[ \sum_{j=1}^n \mid
\chi^j \rangle \langle \chi^j \mid \right] [R(l)Z^l]^\dagger
W(|Z|) {} \\
& & {} \hspace{5cm} \otimes \mid \phi_m \rangle \langle
\phi_l \mid d\mu {} \\
&=& {} \sum_{m=0}^\infty \sum_{l=0}^\infty \int_0^\infty
\int_0^{2\pi} e^{i(m-l)\zeta} N(|Z|)^{-1} [R(m)A(r)^m]
[R(l)A(r)^l]^\dagger W(|Z|) {}\\ 
& & {} \hspace{5cm} \otimes \mid \phi_m \rangle \langle \phi_l
\mid d\mu {}
\end{eqnarray*}
\begin{eqnarray*} 
&=& {} \sum_{m=0}^\infty \left[ 2\pi \int_0^\infty N(|Z|)^{-1}
[R(m)A(r)^m] [R(m)A(r)^m]^\dagger W(|Z|)d\nu \right]  \otimes
\mid \phi_m \rangle \langle \phi_m \mid {} \\
&=& {} \mathbb{I}_n \otimes \sum_{m=0}^\infty \mid \phi_m \rangle
\langle \phi_m \mid = \mathbb{I}_n \otimes I, {}
 \end{eqnarray*}
where we have used the following facts:
$$
\sum_{j=1}^n \mid \chi^j \rangle \langle \chi^j \mid =
\mathbb{I}_n \quad,\quad \int_0^{2\pi} e^{i(m-l)\zeta} d\zeta =
\left\{ \begin{array}{ccc}
 0 & {\text{if}} & m \neq l \\ 2\pi & {\text{if}} & m=l
\end{array} \right.,
$$
and the condition (\ref{ee6}).
 \end{proof}
%-------
Before moving to examples let us make a comment. In general, the $\rho(m)$'s of (\ref{e1}) form a positive sequence
of real numbers. In the examples exhibited hereafter, some of the
entries of the matrix $R(m)$ contain negative values. These
values do not violate the basic definition of moment problems as long as we recover the classical schemes of moment problems. 
%-------
 \begin{example}
 \label{exx1}
Let $x$ be a fixed real number, $r \in [0,\infty)$, and $\zeta \in
[0,2\pi)$. Set
 \begin{equation}
Z = \left( \begin{array}{cc}
 \cos{x} & -\sin{x} \\ \sin{x} & \cos{x}
 \end{array} \right)
 \left( \begin{array}{cc}
\lambda(r) & 0 \\ 0 & \mu(r)
 \end{array} \right)
 \left( \begin{array}{cc}
\cos{x} & -\sin{x} \\ \sin{x} & \cos{x}
 \end{array} \right)^T e^{i\zeta},
 \end{equation}
and
 \begin{equation}
R(m) = \left( \begin{array}{cc}
 \rho_1(m) \cot{x} & \rho_1(m) \\
 \rho_2(m) & -\rho_2(m) \cot{x}
\end{array} \right).
 \end{equation}
Then,
$$
R(m)Z^m = e^{im\zeta} \left( \begin{array}{cc}
\rho_1(m) \lambda(r)^m \cot{x} & \rho_1(m) \lambda(r)^m \\
\rho_2(m) \mu(r)^m & -\rho_2(m) \mu(r)^m \cot{x}
 \end{array} \right),
$$
and
$$
[R(m)Z^m] [R(m)Z^m]^\dagger = \left( \begin{array}{cc}
\rho_1(m)^2 \lambda(r)^{2m} \csc^2{x} & 0 \\
0 & \rho_2(m)^2 \mu(r)^{2m} \csc^2{x}
 \end{array} \right).
$$
Because of the properties of the trace, there is no need to
compute $[R(m)Z^m]^\dagger [R(m)Z^m]$ before knowing its trace,
since, even though the two matrices are different, they have the
same trace. Hence,
$$
\text{Tr} \{ [R(m)Z^m]^\dagger [R(m)Z^m] \} = \csc^2{x} \left[
\rho_1(m)^2 \lambda(r)^{2m} + \rho_2(m)^2 \mu(r)^{2m} \right].
$$
Thus, the normalization condition (\ref{e5}) and the condition for
a resolution of the identity (\ref{ee6}) demand the following
 \begin{equation}
N(|Z|) = \csc^2{x} \sum_{m=0}^\infty \left[ \rho_1(m)^2
\lambda(r)^{2m} + \rho_2(m)^2 \mu(r)^{2m} \right] < \infty, \label{e6}
\end{equation}
and
\begin{equation}
2\pi \left( \begin{array}{cc}
 I_{1} & 0 \\ 0 & I_2
\end{array} \right) = \I_2, \label{e7}
 \end{equation}
where
 \begin{eqnarray*}
I_1 &=& \int_0^\infty N(|Z|)^{-1} \rho_1(m)^2 \lambda(r)^{2m}
\csc^2{x} d\nu, \\
I_2 &=& \int_0^\infty N(|Z|)^{-1} \rho_2(m)^2 \mu(r)^{2m}
\csc^2{x} d\nu.
 \end{eqnarray*}
Let us solve this problem for some special values.
 \begin{itemize}
 \item[(a)] \label{exx1-a} Fix $\displaystyle x=\frac{\pi}{4}$,
$\lambda(r)=r$, $\mu(r)=2r$, $\displaystyle \rho_1(m) =
\frac{1}{\sqrt{m!}}$, and $\displaystyle \rho_2(m) =
\frac{1}{\sqrt{4^m m!}}$. Further, fix the measure as
$\displaystyle d\nu = \frac{2}{\pi }rdr$. Then, (\ref{e6}) and
(\ref{e7}) take the form
 \begin{eqnarray}
N(|Z|) &=& 4 \sum_{m=0}^\infty \frac{r^{2m}}{m!} = 4 e^{r^2},\label{xx} \\
2\pi I_1 &=& 2\pi I_{2} = \frac{\pi}{m!} \int_0^\infty e^{-r^2}
r^{2m} \frac{2}{\pi} rdr = 1.\label{xxx}
 \end{eqnarray}
 \item[(b)] \label{exx1-b} For $\displaystyle x=\frac{\pi}{6}$,
$\lambda(r)=3r$, $\mu(r)=2r$, $\displaystyle \rho_1(m) =
\frac{1}{\sqrt{9^m m!}}$, and $\displaystyle \rho_2(m) =
\frac{1}{\sqrt{4^m m!}}$, let us take the measure to be
$\displaystyle d\nu=\frac{2}{\pi}rdr$. Then, (\ref{e6}) and
(\ref{e7}) become (\ref{xx}) and (\ref{xxx}).
 \end{itemize}
 \end{example}
%---------------------
Now, let us look at a more systematic way of building examples.
\subsection{A particular class of VCS with the R-Z ordering}
\label{g-examples}

Let $B$ be an $n \times n$ fixed matrix such that $BB^T = B^T B =
\I_n$. Let $D = \text{diag}(f_1(z_1),\ldots,f_n(z_n))$, where $z_j
= r_j e^{i\zeta_j}$, $r_j \in \D_j$ (the domain of $r_j$), and
$\zeta_j \in [0,2\pi)$. Form
 \begin{equation}
Z = BDB^T.
 \end{equation}
Then,
$$
Z^m = B D^m B^T.
$$
Let
 \begin{equation}
R(m) = (\rho_1(m) C_1, \rho_2(m) C_2, \ldots, \rho_n(m) C_n)^T,
 \end{equation}
where the $C_j$'s are the columns of $B$. We intend to have VCS
as,
 \begin{equation}
 \label{state1}
\mid Z,j \rangle = N(|Z|)^{-1/2} \sum_{m=0}^\infty R(m)Z^m \chi^j
\otimes \phi_m.
 \end{equation}
Since
 \begin{eqnarray*}
\lefteqn{ [R(m)Z^m] [R(l)Z^l]^\dagger = {} } \\
& & {} = \text{diag} \left( \rho_1(m) \rho_1(l) f_1(z_1)^m
\overline{f_1(z_1)}^l, \ldots, \rho_n(m) \rho_n(l) f_n(z_n)^m
\overline{f_n(z_n)}^l \right), {}
 \end{eqnarray*}
we have
 \begin{eqnarray*}
\lefteqn{ \text{Tr} \{ [R(m)Z^m)]^\dagger [R(m)Z^m] \} =
\sum_{i=1}^n \rho_i(m)^2 |f_i(z_i)|^{2m} {} } \\
& & {} = \rho_1(m)^2 |f_1(z_1)|^{2m} + \rho_2(m)^2 |f_2(z_2)|^{2m}
+ \ldots + \rho_n(m)^2 |f_n(z_n)|^{2m}, {}
 \end{eqnarray*}
and the normalization condition becomes
 \begin{equation}
 \label{nnn1}
N(|Z|) = \sum_{m=0}^\infty \sum_{i=1}^n \rho_i(m)^2
|f_i(z_i)|^{2m} .
 \end{equation}
Setting then
 \begin{equation}
d\mu = d\nu(\zeta_1, \ldots, \zeta_n) d\lambda(r_1, \ldots, r_n),
 \end{equation}
with
 \begin{eqnarray}
d\lambda(r_1, \ldots, r_n) &=& N(|Z|) W(r_1, r_2, \ldots, r_n)
dr_1 dr_2 \ldots dr_n, \\
d\nu(\zeta_1, \ldots, \zeta_n) &=& \frac{d\zeta_1 \ldots
d\zeta_n}{\pi^n},
 \end{eqnarray}
and assuming that
 \begin{equation}
\int_{\D_1} \ldots \int_{\D_n} \int_0^{2\pi} \ldots \int_0^{2\pi}
f_{k_0}(z_{k_0})^m \overline{f_{k_0}(z_{k_0})}^l d\mu = \left\{
\begin{array}{ccc}
0 & \text{if} & m \neq l \\ \pi^n \rho_{k_0}(m) & \text{if} & m=l
\end{array} \right.
 \end{equation}
for some $k_0 \in \{1,2,\ldots,n\}$,
and
$$
\int_{\D_1} \ldots \int_{\D_n} \int_0^{2\pi} \ldots \int_0^{2\pi}
\abs{f_{k}(z_{k})}^{2m} d\mu = \pi^n \rho_k(m)
$$
for all $k \in \{1,2,\ldots,n\}-\{k_0\}$,
if the series in (\ref{nnn1}) converges, then the states in
(\ref{state1}) form a set of VCS. Illustrative examples can easily be seen.
%----------------------------------------------------
\section{VCS with the Z-R ordering}
\label{z-r-order}

So far, we had the matrix $R(m)$ on the left of $Z^m$. If we
change the ordering the construction fails in most of the cases
developed above. In this section, we show that the construction
can be however carried out in the Z-R ordering, that is, when
$R(m)$ is on the right of $Z^m$.
Let us give a simple way of getting VCS of this sort. Consider
$R(m)$ and $Z$ such that
 \begin{eqnarray}
Z &=& B e^{i\theta}, \label{z-in-z-r-ordering} \\
R(m)R(m)^\dagger &=& R(m)^\dagger R(m) = \rho(m) \mathbb{I}_n,
\quad \text{and}\label{cond-r-m-cli}\\ B^m B^{m\dagger} &=& B^{m\dagger} B^m =
f(|Z|)^m \mathbb{I}_n. \label{z-r-cond-on-r-z}
 \end{eqnarray}
For instance, Clifford type matrices satisfy (\ref{z-r-cond-on-r-z}) \cite{TH}. Therefore, we can construct VCS as
 \begin{equation}\label{vcszr}
\mid Z,j \rangle = N(|Z|)^{-1/2} \sum_{m=0}^\infty Z^m R(m) \chi^j
\otimes \phi_m, \quad j=1,2,\ldots,n,
 \end{equation}
provided that
 \begin{equation}
 \sum_{m=0}^\infty f(|Z|)^m\rho(m) < \infty, \quad \text{and} \quad
\int_\mathcal{R} f(|Z|)^m d\mu = \rho(m),
 \end{equation}
where $\mathcal{R}$ is the parametrization domain of $B$, and
$d\mu$ a measure on it. An illustrative example can easily be seen.
%--------------
\section{The generalized oscillator algebra}
\label{algebra}

We aim in this section to define a generalized oscillator algebra
related to the Z-R ordering in the construction of VCS. To this
end, we define a generalized factorial with matrices, and,
thereby, we define a generalized oscillator algebra for the states
in (\ref{vcszr}). Let us recall first how this construction was
done for the states (\ref{e1}).

Let
 \begin{equation}
x_m = \frac{\rho(m)}{\rho(m-1)}, \quad \text{for} \quad m \in
\mathbb{N}^*.
 \end{equation}
Then, the so-called generalized factorial can be defined as
 \begin{equation}
\rho(m) = x_m x_{m-1} \ldots x_1 = x_m!, \quad \forall ~m \in
\N^*,
 \end{equation}
with $x_0!:=1$. For an orthonormal basis $\{\phi_m\}_{m=0}^\infty$
of the Hilbert space $\H$, the generalized annihilation, creation,
and number operators are defined respectively as (see \cite{AAG})
 \begin{eqnarray*}
\an \phi_m &=& \sqrt{x_m} \phi_{m-1}, \quad \text{with} \quad
\an \phi_0 = 0, \\
\an^\dagger \phi_m &=& \sqrt{x_{m+1}} \phi_{m+1}, \\
\en \phi_m &=& x_m \phi_m,
 \end{eqnarray*}
and the commutators take the form
 \begin{eqnarray*}
\left[ \an, \an^\dagger \right] \phi_m &=& (x_{m+1}-x_m) \; I \phi_m, \\
\left[ \en, \an \right] \phi_m &=&  -(x_m-x_{m-1}) \; \an \phi_m, \\
\left[ \en, \an^\dagger \right] \phi_m &=&  (x_{m+1}-x_m) \;
\an^\dagger \phi_m.
 \end{eqnarray*}
The CS, $\mid z \rangle$ are eigenvectors of the annihilation
operator $\an$, that is, $\an \mid z \rangle = z \mid z \rangle$.
Under the commutation operation, these three operators generate a
Lie algebra which is called the {\em generalized oscillator
algebra}, and denoted by $\mathfrak{U}_{\text{osc}}$. In general,
the dimension of this algebra is not finite. From the commutation
relations, it is obvious that the dimension is completely
depending on the form of $x_m$.

In the same spirit, let us define, for an $n \times n$ matrix
$R(m)$,
 \begin{equation}
x_m = R(m)R(m-1)^{-1}, \quad m \in \N^*.
 \end{equation}
Therefore, we can define
 \begin{equation}
\text{for} \quad m \geq 1, \quad x_m! = x_m x_{m-1} \ldots x_1 = R(m),
\quad \text{and} \quad x_0! = \I_n,
 \end{equation}
where we have assumed that $R(m)$ is invertible for all $m \geq
1$ and $R(0)=\I_n$. Here, the annihilation, creation, and number operators have to
be defined on the basis $\{ \chi^j \otimes \phi_m \}_{m \geq 0, \;
j=1,\ldots,n}$. To this end, let us consider the $n \times n$
elementary matrices $E_{ij}$, $i,j = 1,2,\ldots,n$, which have
each a unit in the $(i,j)$th position and zero elsewhere. Note
that, for $i,j,k,\ell = 1,2,\ldots,n$,
 \begin{eqnarray}
E_{ij} E_{k\ell} &=& \delta_{jk} \; E_{i\ell}, \\
E_k x_m E_\ell &=& (x_m)_{kl} E_{k\ell}, \\
x_m E_{k\ell} &=& \sum_{i=1}^n (x_m)_{ik} E_{i\ell}, \quad
\text{and} \quad E_{k\ell} x_m = \sum_{i=1}^n (x_m)_{\ell i} E_{ki}, \\
E_{k\ell} \chi^j &=& \delta_{\ell j} \; \chi^k,
 \end{eqnarray}
where we have denoted $E_k=E_{kk}$, and $\delta_{kj}$ is the
Kronecker symbol. Since $x_m$ does not depend on $j$, for each
$j$, we can define a set of annihilation, creation and number
operator. Let us denote  them by $A_j$, $A_j^\dagger$, $N_j$,
with
 \begin{equation}
A_j = E_j \otimes \an \quad,\quad A_j^\dagger = E_j \otimes
\an^\dagger \quad,\quad N_j = E_j \otimes \en.
 \end{equation}
The action of these operators on the basis elements of $\WH$
should be understood in the following way: For each
$j,k=1,\ldots,n$, we define
 \begin{eqnarray}
A_k \chi^j \otimes \phi_m &=& x_m^{-1} E_k \chi^j \otimes
\phi_{m-1} = \delta_{kj} \; x_m^{-1} \chi^k \otimes \phi_{m-1}, \\
& & \text{with} \quad A_k \chi^j \otimes \phi_0 = 0, \nonumber \\
A_k^\dagger \chi^j \otimes \phi_m &=& x_{m+1}^{-1} E_k \chi^j
\otimes \phi_{m+1} = \delta_{kj} \; x_{m+1}^{-1} \chi^k
\otimes \phi_{m+1}, \\
N_k \chi^j \otimes \phi_m &=& (x_m^{-1} E_k)^2 \chi^j \otimes
\phi_m = \delta_{kj} \; x_m^{-1} E_k x_m^{-1} \chi^k \otimes
\phi_m \\
&=& \delta_{kj} \; (x_m^{-1})_{kk} x_m^{-1} \chi^k \otimes \phi_m,
\nonumber
 \end{eqnarray}
The commutators take then the the form
 \begin{eqnarray*}
\left[ A_k, A_\ell^\dagger \right] \chi^j \otimes \phi_m &=& \{
\delta_{\ell j} x_{m+1}^{-1} E_k x_{m+1}^{-1} \chi^\ell -
\delta_{kj} x_m^{-1} E_\ell x_m^{-1} \chi^k \} \otimes \phi_m \\
&=& \{ \delta_{\ell j} (x_{m+1}^{-1})_{k\ell} x_{m+1}^{-1} \chi^k
- \delta_{kj} (x_m^{-1})_{\ell k} x_m^{-1} \chi^\ell \} \otimes
\phi_m, \\
\left[ N_k, A_\ell \right] \chi^j \otimes \phi_m &=& \{
\delta_{\ell j} (x_{m-1}^{-1})_{kk} x_{m-1}^{-1} E_k x_m^{-1}
\chi^\ell - \delta_{kj} (x_m^{-1})_{\ell k} x_m^{-1} E_{\ell k}
x_m^{-1} \chi^k \} \otimes \phi_{m-1} \\
&=& \{ \delta_{\ell j} (x_{m-1}^{-1})_{kk} (x_m^{-1})_{k\ell}
x_{m-1}^{-1} \chi^k - \delta_{kj} (x_m^{-1})_{\ell k}
(x_m^{-1})_{kk} x_m^{-1} \chi^\ell \} \otimes \phi_{m-1}, \\
\left[ N_k, A_\ell^\dagger \right] \chi^j \otimes \phi_m &=&
x_{m+1}^{-1} \{ \delta_{\ell j} (x_{m+1}^{-1})_{kk} E_k
x_{m+1}^{-1} \chi^\ell - \delta_{kj} (x_m^{-1})_{\ell k} E_{\ell
k} x_m^{-1} \chi^k \} \otimes \phi_{m+1} \\
&=& x_{m+1}^{-1} \{ \delta_{\ell j} (x_{m+1}^{-1})_{kk}
(x_{m+1}^{-1})_{k\ell} \chi^k - \delta_{kj} (x_m^{-1})_{\ell k}
(x_m^{-1})_{kk} \chi^\ell \} \otimes \phi_{m+1}.
 \end{eqnarray*}
We can therefore define the ``global" annihilation, creation and
number operators $A$, $A^\dagger$, and $N$ on $\WH$ as
 \begin{equation}
A = \sum_{k=1}^n A_k = \mathbb{I}_n \otimes \an, \quad A^\dagger =
\sum_{k=1}^n A_k^\dagger = \mathbb{I}_n \otimes \an^\dagger, \quad
N = \sum_{k=1}^n N_k = \mathbb{I}_n \otimes \en.
 \end{equation}
We have then that
 \begin{eqnarray}
A \chi^j \otimes \phi_m &=& x_m^{-1} \chi^j \otimes \phi_{m-1} =
A_j \chi^j \otimes \phi_m,
\quad \text{with} \quad A \chi^j \otimes \phi_0 = 0, \\
A^\dagger \chi^j \otimes \phi_m &=& x_{m+1}^{-1} \chi^j \otimes
\phi_{m+1} = A_j^\dagger \chi^j \otimes \phi_m, \\
N \chi^j \otimes \phi_m &=& (x_m^{-1})_{jj} \; x_m^{-1} \chi^j
\otimes \phi_m = N_j \chi^j \otimes \phi_m,
 \end{eqnarray}
and the commutators read
 \begin{eqnarray}
\left[ A, A^\dagger \right] \chi^j \otimes \phi_m &=&
(x_{m+1}^{-2} - x_m^{-2}) [\mathbb{I}_n \otimes I] \chi^j \otimes \phi_m, \\
\left[ N, A \right] \chi^j \otimes \phi_m &=& - (x_m^{-2} -
x_{m-1}^{-2}) A \chi^{j} \otimes \phi_m, \\
\left[ N, A^\dagger \right] \chi^j \otimes \phi_m &=& x_{m+1}^{-1}
(x_{m+1}^{-1} - x_m^{-2} x_{m+1}) A^\dagger \chi^j \otimes \phi_m.
 \end{eqnarray}
In order to realize the VCS as eigenstates of the annihilation
operator in the Z-R representation, the states can be written in
terms of $x_m$ as
 \begin{equation}
\mid Z,j \rangle = N(|Z|)^{-1/2} \sum_{m=0}^\infty Z^m R(m) \chi^j
\otimes \phi_m = N(|Z|)^{-1/2} \sum_{m=0}^\infty Z^m x_m! \;
\chi^j \otimes \phi_m,
 \end{equation}
and the action of $A_k$ reads,
 \begin{eqnarray}
A_k \mid Z,j \rangle &=& N(|Z|)^{-1/2} \sum_{m=1}^\infty Z^m x_m!
\; \delta_{kj} \; x_m^{-1} \chi^j \otimes \phi_{m-1} \\
&=& \delta_{kj} \; N(|Z|)^{-1/2} \sum_{m=0}^\infty Z^{m+1}
x_{m+1}! \; x_{m+1}^{-1} \chi^j \otimes \phi_m \nonumber \\
&=& \delta_{kj} Z \mid Z,j \nonumber \rangle.
 \end{eqnarray}
It follows immediately that
 \begin{equation}
A \mid Z,j \rangle = \sum_{k=1}^n A_k \mid Z,j \rangle =
\sum_{k=1}^n \delta_{kj} \; Z \mid Z,j \rangle = Z \mid Z,j
\rangle,
 \end{equation}
that is, the Z-R ordering VCS are eigenstates of the annihilation
operator $A$.

Let us look now at the algebra and the actions of the operators for a particular example.
%---------------------------------------------------
 \begin{example}
In Example \ref{exx1-a}-(a), $R(m)$ had the form
$$
R(m) = \left( \begin{array}{cc}
 \displaystyle \frac{1}{\sqrt{m!}} & \displaystyle
\frac{1}{\sqrt{m!}} \\
 \displaystyle \frac{1}{\sqrt{4^m m!}} & \displaystyle
-\frac{1}{\sqrt{4^m m!}}
\end{array} \right).
$$
When $R(m)^\dagger$ is placed on the right of $Z^m$, that is,
$$
\mid Z,j \rangle = N(|Z|)^{-1/2} \sum_{m=0}^\infty Z^m
R(m)^\dagger \chi^j \otimes \phi_m,
$$
the normalization factor and the resolution of the identity remain
the same (the weight is the same). For this particular $R(m)$,
$x_m$ wears the form
 \begin{equation}
x_m = \frac{1}{4\sqrt{m}} \left( \begin{array}{cc}
 3 & 1 \\ 1 & 3
\end{array} \right).
 \end{equation}
Let
$$C=\frac{1}{2} \left( \begin{array}{cc}
 3 & -1 \\ -1 & 3
\end{array} \right),\quad
D=\frac{1}{2} \left(
\begin{array}{cc}
 5 & -3 \\ -3 & 5
\end{array} \right),\quad
E=\frac{1}{4} \left( \begin{array}{cc}
 3 & 1 \\ 1 & 3
\end{array} \right).$$
Under the action of the operators defined as above and their
commutators, in this particular example, we have
 \begin{eqnarray}
A &=& C \otimes \an, \quad
A^\dagger = C \otimes \an^\dagger, \quad
N = \frac{3}{2} C \otimes \en.
 \end{eqnarray}
In this case, the defining relations of the deformed
oscillator algebra are
 \begin{eqnarray}
\left[ A, A^\dagger \right] &=& D \otimes I, \quad
\left[ N, A \right] = - D A, \quad
\left[ N, A^\dagger \right] = D A^\dagger.
 \end{eqnarray}
If we redefine the operators $A$, $A^\dagger$, $N$ as
 \begin{equation}
\widetilde{A} = E A, \quad \widetilde{A}^\dagger = E A^\dagger, \quad
\widetilde{N} = \widetilde{A}^\dagger \widetilde{A} ,
 \end{equation}
we recover the classical harmonic oscillator algebra, with
 \begin{equation}
\left[ \widetilde{A}, \widetilde{A}^\dagger \right] = \I_2 \otimes
I, \quad \left[ \widetilde{N}, \widetilde{A} \right] =
-\widetilde{A}, \quad \left[ \widetilde{N}, \widetilde{A}^\dagger
\right] = \widetilde{A}^\dagger.
 \end{equation}
 \end{example}
%-----------------------------------------------------------
In the following section we discuss physical applications of VCS.
\section{Physical Examples}\label{phys}
In this section we consider two physical examples. As a first example we construct VCS for a special case of the Jaynes-Cummings model and study some physical properties. In the second example we derive VCS for a conditionally exactly solvable supersymmetric harmonic oscillator. 
\subsection{Example: Jaynes-Cummings model}
Let us consider the well-known Jaynes-Cummings Hamiltonian
\cite{DH,K}. It is diagonalizable, and it describes a two-level
atom interacting with a single mode interaction field. In the
rotating wave approximation, it reads ($\hbar=1$)
 \begin{equation}
 \label{JC}
H_{JC} = \omega(\an^\dagger \an + \frac{1}{2}) \sigma_0 +
\frac{\omega_0}{2} \sigma_3 + \kappa(\an^\dagger \sigma_- + \an
\sigma_+)
 \end{equation}
where $\omega$ is the field mode frequency, $\omega_0$ the atomic
frequency, $\kappa$ a coupling constant, $\sigma_0 = \I_2$ is the
$2 \times 2$ identity matrix, $\sigma_1$, $\sigma_2$, $\sigma_3$
are the Pauli matrices, and
 \begin{equation}
\sigma_+ = \sigma_1 + i\sigma_2, \quad \sigma_- = \sigma_1 -
i\sigma_2.
 \end{equation}
The Hamiltonian (\ref{JC}) and its generalizations have been used to study several
physical problems (for example, atomic interactions with
electromagnetic fields \cite{Kasa,K}, spontaneous emissions in cavity \cite{Kle}, Rabi
oscillations \cite{Fuji}, ions in harmonic traps \cite{Cirac}, and quantum computations \cite{Hug}). In the following we construct VCS for a special case of the $H_{JC}$. This set of VCS can be used to compute physical quantities associated to the problem under consideration. We shall compute expectation values, dispersion, average energy and the signal to quantum noise ratio.

It is known that the Hamiltonian $H_{JC}$ can be diagonalized as
$$O^{\dagger}H_{JC}O=H_D=\left(\begin{array}{cc}
H_{D(+)}&0\\
0&H_{D(-)}\end{array}\right),$$
where $O$ is the diagonalization operator. From the diagonal form the energy eigenvalues can be obtained as
$$E_{n}^{+}=\omega n+\kappa r(n)\;\;\text{and}\;\;E_{n}^{-}=\omega (n+1)-\kappa r(n+1),$$
where $r(n)=\sqrt{\delta+n},~\delta=\left(\frac{\Delta}{2\kappa}\right)^2$ and $\Delta=\omega-\omega_0$ is the detuning with $\Delta>0$. Since $\Delta>0$ we have $E_{n+1}^{-}>E_m^-.$ If $0<\kappa/\omega\leq 2\sqrt{\delta+1}$ the energies $E_n^+$ are strictly increasing and non-degenerate \cite{DH}. Let
$$\omega_{\pm}=\frac{\omega\pm\kappa^2}{\Delta}\;\;\text{and}\;\;e_n^{\pm}=E_n^{\pm}-E_0^{\pm}.$$
In the weak coupling limit case we expand $e_n^{\pm}$ and by keeping at most terms of order two in $\kappa$ we get \cite{DH}
$$e_n^{\pm}(\kappa<<)=\omega_{\pm}(\kappa)n.$$
In this case, let us again denote the diagonalized version of the Hamiltonian $H_{JC}$ by $H_D$ and let $\psi_n^{\pm}$ be the corresponding normalized energy states. Set
$$\rho_{\pm}(n)=e_1^{\pm}e_2^{\pm}\cdots e_n^{\pm}=\left[\omega_{\pm}(\kappa)\right]^n\Gamma(n+1)$$
Since the Hilbert space of $H_{JC}$ can be taken \cite{DH} as the linear span of 
$$\left\{\psi_n^-=\left(\begin{array}{c}0\\\mid n\rangle\end{array}\right),\psi_n^+=\left(\begin{array}{c}\mid n\rangle\\0\end{array}\right):~n=0,1,2,..\right\}$$
we make the following identification
\begin{equation}\label{wave}
\psi_n^{+}:=\left(\begin{array}{cc}\phi_n\\0\end{array}\right)=\chi_1\otimes\phi_n\;\;\;\text{and}\;\;\;
\psi_n^{-}:=\left(\begin{array}{cc}0\\\psi_n\end{array}\right)=\chi_2\otimes\phi_n,
\end{equation}
where $\{\chi_1,\chi_2\}$ is the natural basis of $\mathbb{C}^2$ and $\{\phi_n\}$ is an orthonormal basis of a Hilbert space $\mathfrak{H}$.
Set
$$R(n)=\text{diag}(\rho_+(n),\rho_-(n))\;\;\;\text{and}\;\;\;Z=\text{diag}(z_1,z_2).$$
\begin{equation}\label{jccs}
\mid Z,j\rangle=\mathcal{N}(Z)^{-\frac{1}{2}}\sum_{n=0}^{\infty}R(n)^{-\frac{1}{2}}Z^n\chi_j\otimes\phi_n;~~j=1,2.
\end{equation}
forms a set of CS for the Hamiltonian $H_D$, where $z_1=r_1e^{i\theta_1},z_2=r_2e^{i\theta_2},r_1,r_2\in[0,\infty)$ and $\theta_1,\theta_2\in[0,2\pi)$. In this case, the normalization factor is given by
$$\mathcal{N}(Z)=e^{r_1^2/\omega_+(\kappa)}+e^{r_2^2/\omega_-(\kappa)}$$
 and a resolution of the identity is obtained with the measure
$$d\mu(Z)=\frac{r_1r_2}{\pi^2\omega_+(\kappa)\omega_-(\kappa)}e^{-r_1^2/\omega_+(\kappa)}e^{-r_2^2/\omega_-(\kappa)}\mathcal{N}(Z)dr_1dr_2d\theta_1d\theta_2.$$
Let $c_1,c_2\in\C$ with $\abs{c_1}^2+\abs{c_2}^2=1$ then the vectors
\begin{equation}\label{gen}
\mid Z\rangle=c_1\mid Z,1\rangle+c_2\mid Z,2\rangle
\end{equation}
form a general set of CS for the Hamiltonian $H_{D}$ and these CS can be transformed back to the original Hamiltonian using the unitary operator $O$. Further, under this transformation the mean values are invariant \cite{DH}.
For $x\in\mathbb{R}^+$ let $$U=\left(\begin{array}{cc}\cos{x}&-\sin{x}\\\sin{x}&\cos{x}\end{array}\right).$$
since $U$ is a unitary matrix, in (\ref{jccs}) if we replace $R(n)^{-\frac{1}{2}}Z^n$ by  $UR(n)^{-\frac{1}{2}}Z^nU^{\dagger}$ the resulting vectors still form a set VCS with the same normalization factor as that of (\ref{jccs}). In this case a resolution of the identity is obtained with the measure $d\zeta(Z,U)=d\mu(Z)d\nu(U)$ where $d\nu(U)$ is the normalized invariant measure of $\mathbb{R}^+$. 
Further observe that
\begin{equation}\label{su}
U\mid Z,k\rangle U^{\dagger}\not=\mathcal{N}(Z)^{-\frac{1}{2}}\sum_{n=0}^{\infty}UR(n)^{-\frac{1}{2}}Z^nU^{\dagger}\chi_k\otimes\phi_n=\mid Z,U,k\rangle.
\end{equation}
Let
\begin{eqnarray*}
\mid z_1\rangle&=&\frac{1}{e^{r_1^2/\omega_+(\kappa)}+e^{r_2^2/\omega_-(\kappa)}}\sum_{n=0}^{\infty}\frac{z_1^n}{\sqrt{\omega_+(\kappa)^nn!}}\phi_n\\
\mid z_2\rangle&=&\frac{1}{e^{r_1^2/\omega_+(\kappa)}+e^{r_2^2/\omega_-(\kappa)}}\sum_{n=0}^{\infty}\frac{z_2^n}{\sqrt{\omega_-(\kappa)^nn!}}\phi_n.
\end{eqnarray*}
Then the states $\mid Z,U,k\rangle$ can be explicitly written as
$$\mid Z,U,1\rangle=\left(\begin{array}{c}\cos^2{x}\mid z_1\rangle+\sin^2{x}\mid z_2\rangle\\
\sin{x}\cos{x}(\mid z_1\rangle-\mid z_2\rangle)\end{array}\right),\quad
\mid Z,U,2\rangle=\left(\begin{array}{c}
\sin{x}\cos{x}(\mid z_1\rangle-\mid z_2\rangle)\\
\sin^2{x}\mid z_1\rangle+\cos^2{x}\mid z_2\rangle
\end{array}\right).$$
If we set $x=0$ in $\mid Z,U,k\rangle$ we recover $\mid Z,k\rangle$.
In (\ref{jccs}) if we replace $Z$ and $R(n)^{-\frac{1}{2}}$ respectively by $UZU^{\dagger}$ and $(\sqrt{\rho_+(n)}C_1,
 \sqrt{\rho_-(n)}C_2)^T$, where $C_1$ and $C_2$ are the column vectors of $U$, by the argument of the Subsection \ref{g-examples} we can have a set of VCS associated with the Hamiltonian $H_{D}$. In fact the sets of VCS (\ref{jccs}) and (\ref{su}) can be considered as a blend of standard spin CS and the canonical CS with $n!$ replaced by $\omega_{\pm}(\kappa)^nn!$\cite{P,TA}. Thus these VCS may be considered as a coherent state wavefunction of a non-relativistic two-level particle in the weak coupling limit.
In the weak coupling limit case, the Hamiltonian $H_D$ can be written as
\begin{eqnarray*}
H_D&=&\left(\begin{array}{cc}H_{D(+)}&0\\0&H_{D(-)}\end{array}\right) 
=\left(\begin{array}{cc}\omega_+a^{\dagger}a&0\\0&\omega_-a^{\dagger}a\end{array}\right)\\
&=& \left(\begin{array}{cc}\sqrt{\omega_+}a^{\dagger}&0\\0&\sqrt{\omega_-}a^{\dagger}\end{array}\right)
\left(\begin{array}{cc}\sqrt{\omega_+}a&0\\0&\sqrt{\omega_-}a\end{array}\right)
=A^{\dagger}A,
\end{eqnarray*}
where $a\phi_n=\sqrt{n}\phi_{n-1}$ and $a^{\dagger}\phi_n=\sqrt{n+1}\phi_{n+1}$. $A$ and $A^{\dagger}$ are the annihilation and creation operators for $H_D$. Let $N=A^{\dagger}A=H_D$. We can also define the self-adjoint quadrature operators:
$$Q=\frac{A+A^{\dagger}}{\sqrt{2}},\;\;\;\;P=\frac{A-A^{\dagger}}{i\sqrt{2}}.$$
 Under the interpretation that the constructed VCS are the coherent state wavefunction of a two-level particle, the following mean values can be interpreted as the average values of the observables that we would expect to obtain for the particle from a large number of measurements. The mean value of an operator $F$ is a state $\psi$ is given by $\langle F\rangle_{\psi}=\langle\psi\mid F\mid \psi\rangle.$ 
Let
$$G(r_1,r_2)=\frac{e^{r_1^2/\omega_+}}{e^{r_1^2/\omega_+}+e^{r_2^2/\omega_-}},\quad {\mathfrak{G}}(r_1,r_2)=\frac{e^{r_2^2/\omega_-}}{e^{r_1^2/\omega_+}+e^{r_2^2/\omega_-}}.$$
Let us see the mean values of the operators associated with $H_D$ in the states (\ref{jccs}).
\begin{eqnarray*}
&&\langle A\rangle_{\mid Z,1\rangle}=z_1\g,\quad
\langle A\rangle_{\mid Z,2\rangle}=z_2\G
\quad \langle A^{\dagger}\rangle_{\mid Z,1\rangle}=\overline{z}_1\g,\\
&&\langle A^{\dagger}\rangle_{\mid Z,2\rangle}=\overline{z}_2\G,
\quad \langle H_D\rangle_{\mid Z,1\rangle}=r_1^2\g,\quad
\langle H_D\rangle_{\mid Z,2\rangle}=r_2^2\G\\
&&\langle Q\rangle_{\mid Z,1\rangle}=\sqrt{2}r_1\cos{\theta_1}\g,\quad
\langle Q\rangle_{\mid Z,2\rangle}=\sqrt{2}r_2\cos{\theta_2}\G\\
&&\langle P\rangle_{\mid Z,1\rangle}=\sqrt{2}r_1\sin{\theta_1}\g,\quad
\langle P\rangle_{\mid Z,2\rangle}=\sqrt{2}r_2\sin{\theta_2}\G
\end{eqnarray*}
Since $\g,\G<1$, the above mean values are the truncated version of the ordinary harmonic oscillator mean values. Further, since the vectors $\chi_1\otimes\phi_n$ and $\chi_2\otimes\phi_n$ are orthogonal the mean values of the operators in the general set of VCS (\ref{gen}) can be directly obtained from the calculated mean values. For example,
\begin{equation}\label{genmean}
\langle Q\rangle_{\mid Z\rangle}=\abs{c_1}^2\langle Q\rangle_{\mid Z,1\rangle}+\abs{c_2}^2\langle Q\rangle_{\mid Z,2\rangle}.
\end{equation}
The mean value of $H_D$ in the states $\mid Z,U,k\rangle$ of (\ref{su}) takes the form
\begin{eqnarray*}
\langle H_D\rangle_{\mid Z,U,1\rangle}&=&r_1^2\cos^2{x}\g+\frac{r_2^2\omega_+}{\omega_-}\sin^2{x}\G\\
\langle H_D\rangle_{\mid Z,U,2\rangle}&=&\frac{r_1^2\omega_-}{\omega_+}\sin^2{x}\g+r_2^2\cos^2{x}\G,
\end{eqnarray*}
which are the average energies of the particle, in the weak coupling limit, in the coherent state wavefunctions $\mid Z,U,k\rangle.$ Here again one can write a general set of VCS
\begin{equation}\label{ugen}
\mid Z,U\rangle=c'_1\mid Z,U,1\rangle+c'_2\mid Z,U,2\rangle,
\end{equation}
where $c'_1,c'_2\in\mathbb{C}$ with $\abs{c'_1}^2+\abs{c'_2}^2=1$, and obtain the mean value of $H_D$ in the general set of VCS using a relation similar to (\ref{genmean}). Roughly speaking, the dispersion of an observable characterizes ``fuzziness" \cite{Sa}. The dispersion of an operator $F$ in a state $\mid\psi\rangle$ is given by
$$(\Delta F)^2_{\mid\psi\rangle}=\langle\psi\mid F^2\mid\psi\rangle-\langle\psi\mid F\mid\psi\rangle^2.$$
In order to obtain the dispersion, first we calculate the mean values of $H_D^2$.
\begin{eqnarray*}
&&\langle H_D^2\rangle_{\mid Z,1\rangle}=r_1^2(r_1^2+\omega_+)\g,\quad 
\langle H_D^2\rangle_{\mid Z,2\rangle}=r_2^2(r_2^2+\omega_-)\G,
\end{eqnarray*}
For the states (\ref{su}) we have
\begin{eqnarray*}
&&\langle H_D^2\rangle_{\mid Z,U,1\rangle}=r_1^2(r_1^2+\omega_+)\cos^2{x}\g+\frac{\omega_+^2r_2^2(r_2^2+\omega_-)}{\omega_-^2}\sin^2{x}\G,\\ 
&&\langle H_D^2\rangle_{\mid Z,U,2\rangle}=\frac{\omega_-^2r_1^2(r_1^2+\omega_+)}{\omega_+^2}\sin^2{x}\g+r_2^2(r_2^2+\omega_-)\cos^2{x}\G
\end{eqnarray*}
For the general sets of states (\ref{gen}) and (\ref{ugen}), the mean value of $H^2_D$ can be obtained using a relation similar to (\ref{genmean}). The dispersion of $H_D$ in different sets of VCS is now straightforward. In the same manner we obtain the dispersion of $P$ and $Q$ as follows:
\begin{eqnarray*}
\left(\Delta Q\right)^2_{\mid Z,1\rangle}&=&2r_1^2\cos^2{\theta_1}\g\G+\frac{\omega_+}{2}\g\\
\left(\Delta Q\right)^2_{\mid Z,2\rangle}&=&2r_2^2\cos^2{\theta_2}\g\G+\frac{\omega_-}{2}\G\\
\left(\Delta P\right)^2_{\mid Z,1\rangle}&=&2r_1^2\sin^2{\theta_1}\g\G+\frac{\omega_+}{2}\g\\
\left(\Delta P\right)^2_{\mid Z,2\rangle}&=&2r_2^2\sin^2{\theta_2}\g\G+\frac{\omega_-}{2}\G.
\end{eqnarray*}
Thereby one can obtain the uncertainty product $\left(\Delta Q\right)^2_{\mid Z,1\rangle}\left(\Delta P\right)^2_{\mid Z,1\rangle}$ in a straightforward way. Noise is, loosely, any disturbance tending to interfere with the normal operation of a system. For a state $\mid\psi\rangle$ the signal-to-quantum-noise ratio (SNR) is defined as
$$\sigma_{\mid\psi\rangle}=\frac{\langle Q\rangle_{\mid\psi\rangle}}{(\Delta Q)^2_{{\mid\psi\rangle}}}.$$
A high SNR indicates that the noise dominate the measurement, a low SNR indicates a relatively clean measurement. The SNR for various VCS can be seen readily and thereby the noise associated with the measurements can be observed. For example,
$$\sigma_{\mid Z,1\rangle}=\frac{2r_1^2\cos^2{\theta_1}\g^2}{4r_1^2\cos^2{\theta_1}\g\G+\omega_-\G}.$$
Let $Z_{\pm}(t)=Ze^{-i\omega_{\pm}t}$. The time evolution operator of $H_D$ takes the form
$$T(t)=e^{-iH_Dt}=\text{diag}\left(e^{-iH_{D(+)}t},e^{-iH_{D(-)}t}\right).$$
Since $T(t)\chi_1\otimes\phi_n=e^{-i\omega_+nt}\chi_1\otimes\phi_n$ and $T(t)\chi_2\otimes\phi_n=e^{-i\omega_-nt}\chi_2\otimes\phi_n$ we have $T(t)\mid Z,1\rangle=\mid Z_+(t),1\rangle$ and $T(t)\mid Z,2\rangle=\mid Z_-(t),2\rangle$. Thus the VCS $\mid Z,k\rangle$ are temporally stable. Similarly the time evolution of $\mid Z,U,1\rangle,\mid Z,U,2\rangle$ and the general sets of VCS can be seen. In terms of the state $\mid\psi\rangle$, the so-called Mandel parameter is given by
$$Q_{\mid\psi\rangle}^M=\frac{\left(\Delta H_D\right)^2_{\mid\psi\rangle}}{\langle H_D\rangle_{\mid\psi\rangle}}-1.$$
Here again it is straightforward to calculate the Mandel parameter for the classes of VCS discussed above. For example,
$$Q^M_{\mid Z,1\rangle}=r_1^2\G+\omega_+-1.$$
%---------------------------------------------------------
\subsection{The radial harmonic oscillator with unbroken SUSY}\label{susy}
For the sake of completeness, first we briefly introduce the radial harmonic oscillator with unbroken SUSY (for short RHO).
In the supersymmetric setup the SUSY Hamiltonian can be written as
$$H=\left(\begin{array}{cc}H_+&0\\0&H_-\end{array}\right),$$
where (units are such that $\hbar=m=1$)
$$H_{\pm}=-\frac{1}{2}\frac{d^2}{dx^2}+V_{\pm}(x),\;\;V_{\pm}(x)=\frac{1}{2}(W^2(x)\pm W'(x))$$
and $W:M\longrightarrow\mathbb{R}$ is the SUSY potential with $M$ being the configuration space. The SUSY partner Hamiltonians can be written as
$$H_+=AA^{\dagger}\geq 0,\;\;\;H_-=A^{\dagger}A\geq 0,$$
where $A$ and $A^{\dagger}$ are the supercharge operators:
$$A=\frac{1}{\sqrt{2}}\left(\frac{d}{dx}+W(x)\right),\;\;\;A^{\dagger}=\frac{1}{\sqrt{2}}\left(-\frac{d}{dx}+W(x)\right).$$
Since $AH_-=H_+A$ and $H_-A^{\dagger}=A^{\dagger}H_+$ the Hamiltonians $H_+$ and $H_-$ are essentially isospectral \cite{JR1,JR2}. However, there may exist an additional vanishing eigenvalue for one of these Hamiltonians. In this case SUSY is said to be unbroken and by convention this additional eigenvalue is assumed to belong to $H_-$. For the unbroken SUSY the situation can be summarized as follows:
$$H_{\pm}\psi_n^{\pm}=E_{n}^{\pm}\psi_n^{\pm},\;\;n=0,1,2,...$$
where
\begin{eqnarray*}
E_0^-&=&0,\;\;\;\;\;\psi_0^-(x)=C\exp\left(-\int W(x)dx\right)\\
E_{n+1}^{-}&=&E_n^{+}>0,\;\;\;\;\;\psi_{n+1}^-(x)=\sqrt{E_n^+}A^{\dagger}\psi_n^+(x)\\
\psi_n^+(x)&=&\sqrt{E_{n+1}^-}A\psi_{n+1}^-(x).
\end{eqnarray*}
and $C$ is a normalization constant. We consider a conditionally exactly solvable RHO with $M=\mathbb{R}^+$ and
\begin{eqnarray*}
V_+(x)&=&\frac{x^2}{2}+\frac{(\gamma+1)(\gamma+1)}{2x^2}+\varepsilon-\gamma-\frac{3}{2},\\
V_-(x)&=&\frac{x^2}{2}+\frac{\gamma(\gamma+2)}{2x^2}-\varepsilon-\gamma-\frac{1}{2}+\frac{u'(x)}{u(x)}\left(2x-2\frac{\gamma+1}{x}+\frac{u'(x)}{u(x)}\right),
\end{eqnarray*}
where $\gamma\geq 0$, $\varepsilon >-1$ and
$$u(x)={}_1F_1(\frac{1-\varepsilon}{2},-\gamma-\frac{1}{2}-x^2)+\beta x^{2\gamma+3}{}_1F_1(2+\gamma-\frac{\varepsilon}{2},\frac{5}{2}+\gamma,-x^2).$$
Further, the positivity of the solutions requires the following conditions on the parameters (for details see \cite{JR2}):
$$0< \frac{\Gamma(-\gamma-\frac{1}{2})}{\Gamma(\varepsilon/2-\gamma-1)},\quad \abs{\beta}<\frac{\Gamma(-\gamma-\frac{1}{2})\Gamma(\frac{1+\varepsilon}{2})}{\Gamma(\varepsilon/2-\gamma-1)\Gamma(5/2+\gamma)}.$$
With these potentials, as SUSY remain unbroken for all the allowed values of the parameters we have
$$H_{\pm}\psi_n^{\pm}=E_n^{\pm}\psi_n^{\pm}$$
where
$$E_{n+1}^-=E_n^+=2n+1+\varepsilon,\;\;E_0^-=0,\;\;n=0,1,2,...$$
and $\psi_n^{\pm}$ are the normalized energy states. The explicit expressions of $\psi_n^{\pm}$ and further details on the RHO can be found in \cite{JR1,JR2}. Let $e_n^+=E_n^+-E_0^+=2n$, thereby we have
$$0=e_0^+<e_1^+<\cdots<e_n^+<\cdots.$$
Since $E_0^-=0$ and $E_n^-$ strictly increasing, we do not have to shift the spectrum backward. Let
\begin{eqnarray*}
\rho_+(n)&=&e_1^+e_2^+\cdots e_n^+=2^n\Gamma(n+1)\\
\rho_-(n)&=&E_1^-E_2^-\cdots E_n^-=2^n\left(\frac{\varepsilon+3}{2}\right)_n.
\end{eqnarray*}
where $(a)_n=\Gamma(n+a)/\Gamma(a)$ is the Pochhammer symbol. Let us identify the energy states $\psi_n^{\pm}$ to $\chi_j\otimes\phi_n,~j=1,2$ as stated in (\ref{wave}) and take
$$R(n)=\text{diag}(\rho_+(n),\rho_-(n)),\;\;\;Z=\text{diag}(z_1,z_2),$$
where $z_1,z_2$ are as in the previous example. With the above setup the set of vectors
\begin{equation}\label{rho}
\mid Z,j\rangle=\mathcal{N}(Z)^{-\frac{1}{2}}\sum_{n=0}^{\infty}R(n)^{-\frac{1}{2}}Z^n\chi_j\otimes\phi_n,\;\;j=1,2
\end{equation} 
forms a set of CS for the RHO, where
$$\mathcal{N}(Z)=e^{r_1^2/2}+{}_1F_1(1,\frac{\varepsilon+3}{2},\frac{r_2^2}{2})>0,$$
which is finite for all $r_1,r_2>0$. A resolution of the identity is obtained with the measure
$$d\mu(Z)=\mathcal{N}(Z)\frac{r_1r_2}{\pi^2 2^{\frac{\varepsilon+3}{2}}\Gamma(\frac{\varepsilon+3}{2})}e^{-r_1^2/2}e^{-r_2^2/2}dr_1dr_2d\theta_1 d\theta_2.$$
%---------------------------------------------------------
\section{remarks and discussion}
In the case of broken SUSY $H_+$ and $H_-$ are strictly isospectral. The eigenvalues and eigenfunctions are related as follows:
\begin{eqnarray*}
E_n^-&=&E_n^+>0\\
\psi_n^-(x)&=&\sqrt{E_n^+}A^{\dagger}\psi_n^+(x)\\
\psi_n^+(x)&=&\sqrt{E_n^-}A\psi_n^-(x)
\end{eqnarray*}
In this case, it may be interesting to note that the quaternionic VCS discussed in \cite{TA,AEG} can be realized as CS of the supersymmetric Hamiltonian
$$H=\left(\begin{array}{cc}H_+&0\\0&H_-\end{array}\right)$$
For this, let
$$e_n^+=e_n^-=E_n^+-E_0^+=E_n^--E_0^-.$$
Assume that
$$0=e_0^+=e_0^-<e_1^+=e_1^-<\cdots<e_n^+=e_n^-<\cdots.$$
Note that the radial harmonic oscillator with broken SUSY given in \cite{JR2} satisfies this requirement. Let
$\rho(n)=e_n^+!=e_n^-!$ and $Z=\text{diag}(z_1,z_2)$. Identify the wavefunctions $\psi_n^{\pm}$ to $\chi_j\otimes\phi_n$ as before. The set of vectors
\begin{equation}\label{broke}
\mid Z,j\rangle=\mathcal{N}(Z)^{-\frac{1}{2}}\sum_{n=0}^{\infty}\frac{Z^n}{\sqrt{\rho(n)}}\chi_j\otimes\psi_{n},\quad j=1,2
\end{equation}
forms a set of CS for the Hamiltonian $H$. Let $U\in SU(2)$ then 
\begin{equation}\label{broke2}
\mid Z,j\rangle=\mathcal{N}(Z)^{-\frac{1}{2}}\sum_{n=0}^{\infty}\frac{\left(UZU^{\dagger}\right)^n}{\sqrt{\rho(n)}}\chi_j\otimes\psi_{n},\quad j=1,2
\end{equation}
form a set of CS with the same normalization constant of (\ref{broke}). In (\ref{broke}) if a resolution of the identity is obtained with the measure $d\mu(Z)$ then $d\mu(Z)d\nu(U)$ produce a resolution of the identity for the states (\ref{broke2}), where $d\nu(U)$ is the normalized invariant measure of $SU(2)$ \cite{AEG}. If $z_1=z$, $z_2=\overline{z}$ and $e_n^+=e_n^-=n$ then we obtain the quaternionic VCS discussed in \cite{TA} (for a detailed explanation see \cite{AEG}). In such a case, under the setup presented in \cite{TA}, the states (\ref{broke2}) satisfy the three equivalent definitions of the harmonic oscillator canonical CS (for details see \cite{TA}). However, since the partner Hamiltonians $H_+$ and $H_-$ do not posses the same ladder operators \cite{JR1,JR2} for the Hamiltonian $H$ the same properties may not be achieved in the present setup. In the literature, supercoherent states have been studied for a long time \cite{Jay,Or,Fat,Ara}. Main attention has been paid on the supersymmetric linear harmonic oscillator. However, the methods used in the literature were different from the above setup. In most cases, supercoherent states were derived as eigenstates of a supersymmetric annihilation operator \cite{Ara,Fat}. For the supersymmetric harmonic oscillator they were also realized as the minimum uncertainty states with certain exceptions \cite{Ara}. In \cite{Fat} using a supergroup, supercoherent states were derived with a displacement operator. In a future work, for various supersymmetric Hamiltonians we shall study these features in detail under the VCS setup.
%%%%%%%%%%%%%%%%
\section*{Acknowledgments}
The authors are grateful to one of the referees for his valuable comments.


\begin{thebibliography}{XXXX}
\bibitem{AAG} Ali, S.T., Antoine, J-P., Gazeau, J-P.,
{\em Coherent States, Wavelets and their Generalizations},
Springer-Verlag, New York (2000).
\bibitem{KS} Klauder J.R, Skagerstam B.S, {\em Coherent States,
Applications in Physics and Mathematical Physics}, World
Scientific, Singapore, (1985).
\bibitem{P} P\'er\'elomov, A.M., {\em Generalized Coherent States
and Their Applications}, Springer-Verlag, Berlin, (1986).
\bibitem{B} Borzov, V.V., {\em Integral Transforms and
Special Functions} {\bf 12} (2001) 115-138.
\bibitem{O} Odzijewicz, A., {\em
Commun. Math. Phys.} {\bf 192} (1998) 183-215.
\bibitem{BG} Barut, A.O., Girardello, L., {\em Commun. Math. Phys.} {\bf 21} (1971) 41-55.\
\bibitem{Coo} Cooper, L.I., {\em J. Phys. A: Math. Gen.} {\bf 26} (1993) 1601-1623.
\bibitem{Po} Popov, D., {\em J. Phys. A: Math. Gen.} {\bf 34} (2001) 5283-5296.
\bibitem{KPS} Klauder, J.R., Penson, K., Sixdeniers, J-M., {\em Phys. Rev. A} {\bf 64} 013817.
\bibitem{NS} Nieto, M.M., Simmons, L.M., {\em Phys. Rev. D.} {\bf 20} (1979) 1321-1331.
\bibitem{GK} Gazeau, J-P., Klauder, J.R., {\em J. Phys. A: Math. Gen.} {\bf 32} (1999) 123-132.
\bibitem{NG} Novaes, M., Gazeau, J-P., {\em J. Phys. A: Math. Gen.} {\bf 36} (2003) 199-212.
\bibitem{TA} Thirulogasanthar, K., Ali, S.T., {\em J. Math. Phys.} {\bf 44} (2003) 5070-5083.
\bibitem{AEG} Ali, S.T., Englis, M., Gazeau, J-P., {\em J. Phys. A: Math. Gen.} {\bf 37} (2004) 6067-6089.
\bibitem{RD} Rowe, D.J., Repka, J., {\em J. Math. Phys.} {\bf 32} (1991) 2614-2634.
\bibitem{DH} Daoud, M., Hussin, V., {\em J. Phys. A: Math.
Gen.} {\bf 35} (2002) 7381-7402.
\bibitem{Ja} Janowicz, M.W., Ashbourn, J.M.A., {\em Phys. Rev. A} {\bf 55}, (1997) 2348-2359.
\bibitem{As} Ashraf, M.M., {\em Phys. Rev. A} {\bf 50} (1994) 5116-5121.
\bibitem{Gao} Gao, Y.F., Feng, J., Shi, S.R., {\em Internat. J. Theoret. Phys.} {\bf 41} (2002) 867-875.
\bibitem{TH} Thirulogasanthar, K., Hohou\'eto, A.L., Preprint, math-ph/0308020.
\bibitem{K} Kochetov, E.A., {\em J. Phys. A:
Math. Gen.} {\bf 20} (1987) 2433-2442.
\bibitem{Kasa} Kazakov, A.Y., {\em Phys. Lett. A} {\bf 206} (1995) 229-234.
\bibitem{Kle} Kleppner, D., {\em Phys. Rev. Lett.} {\bf 47} (1981) 233-236.
\bibitem{Fuji} Fujii, K., {\em J. Phys. A: Math. Gen.} {\bf 36} (2003) 2109-2124.
\bibitem{Cirac} Cirac, J.I., Blatt, R., Zoller, P., Phillips, W.D., {\em Phys. Rev. A} {\bf 46} (1992) 2668-2681.
\bibitem{Hug} Hughes, R.J. {\em et al.}, {\em Fortsch. Phys.} {\bf 46} (1998) 329-361.
\bibitem{Sa}Sakueai, J.J., {\em Modern Quantum Mechanics}, Addison-Wesley Pub. Co., Reading, Mass., (1994).
\bibitem{PF} Plastina, F., Falci, G., {\em Phys. Rev. B} {\bf 67} (2003) 224514.
\bibitem{AH} Amico, L., Hikami, K., 
Preprint, cond-mat/0309680.
\bibitem{JR1} Junker, G., Roy, P., {\em Phys. Atomic. Nuclei.} {\bf 61} (1998) 1736-1743.
\bibitem{JR2} Junker, G., Roy, P., {\em Phys. Lett. A} {\bf 232} (1997) 155-161.
\bibitem{Jay} Jayaraman, J., de Lima Rodrigues, R., Vaidya, A.N., {\em J. Phys. A: Math. Gen.} {\bf 32} (1999) 6643-6652.
\bibitem{Or} Orszak, M., Salamo, S., {\em J. Phys. A: Math. Gen.} {\bf 21} (1988) L1059-L1064. 
\bibitem{Fat} Fatyga, B.W., Kosteleck\'y, V.A., Nieto, M.M., Truax, D.R., {\em Phys. Rev. D.} {\bf 43} (1991) 1403-1412.
\bibitem{Ara} Aragone, C., Zypman, F., {\em J. Phys. A: Math. Gen.} {\bf 19} (1986) 2267-2279.
\end{thebibliography}
\end{document}